\documentclass[12pt]{iopart}

\usepackage{url}
\usepackage{lineno}
\usepackage{graphicx}  
\newcommand{\pT} {\mbox{$p_{\mathrm{T}}$} }

\begin{document}

\title[ATLAS $v_n$ results]{Measurement of elliptic and higher order flow from ATLAS experiment at the LHC}
\author{Jiangyong Jia for the ATLAS Collaboration} 
\address{Department of Chemistry, Stony Brook University, Stony Brook, NY 11794, USA\\Physics Department, Brookhaven National Laboratory, Upton, NY 11796, USA}
\ead{jjia@bnl.gov}
\begin{abstract}
We present a differential measurement of the azimuthal anisotropy of charged hadron production in Pb+Pb collisions at $\sqrt{s_{\mathrm{NN}}}$=2.76 TeV. This azimuthal anisotropy is expanded into a Fourier series in azimuthal angle, where the coefficient for each term, $v_n$, characterizes the magnitude of the anisotropy at a particular angular scale. We extract $v_2-v_6$ via a discrete Fourier analysis of the two-particle $\Delta\phi-\Delta\eta$ correlation with a large $\Delta\eta$ gap ($|\Delta\eta|>2$), and via an event plane method based on the Forward Calorimeter. Significant $v_2-v_6$ values are observed over a broad range in $\pT$, $\eta$ and centrality, and they are found to be consistent between the two methods in the transverse momentum region $\pT<3-4$ GeV. This suggests that the measured $v_2-v_6$ obtained from two-particle correlations at low $\pT$ with a large $\Delta\eta$ gap are consistent with the collective response of the system to the initial state geometry fluctuations, and is not the result of jet fragmentation or resonance decay.
\end{abstract}
Among the many striking results obtained at RHIC, one important observation is the novel ``ridge''-like and ``cone''-like structures of the two-particle correlation in relative azimuthal angle $\Delta\phi=\phi_{\mathrm{a}}-\phi_{\mathrm{b}}$ and pseudorapidity $\Delta\eta=\eta_{\mathrm{a}}-\eta_{\mathrm{b}}$ in Au+Au collisions~\cite{Adare:2008cqb}. These structures, obsent in elementary proton-proton collisions, are found to extend over a large $\Delta\eta$ range, and show rich $\pT$ and centrality dependence. They were initially interpreted as response of the medium to the energy deposited by the quenched jets~\cite{CasalderreySolana:2004qm}. However, recent studies~\cite{Alver:2010gr} suggest that they could be related to the initial geometric fluctuations and strong collective flow. In this scenario, the spatial fluctuations of nucleons lead to shape deformation at various angular scales, which induce high-order anisotropies of the emitted particles through collective expansion:
\begin{eqnarray}
\label{eq:1}
dN/d\phi \propto 1+2\sum_{n=1}^{\infty}v_n(\pT) \cos \left(n\left(\phi-\Psi_n\right)\right),
\end{eqnarray}
where $v_n$ is the magnitude of the $n^{\mathrm{th}}$ order harmonic flow~\cite{Poskanzer:1998yz}. The two-particle correlation, being a simple convolution of two single particle distribution, is naturally influenced by the same harmonic flow:
\begin{eqnarray}
\label{eq:2}
 dN/d\Delta\phi \propto 1+2\sum_{n=1}^{\infty}v_n(\pT^{\mathrm{a}})v_n(\pT^{\mathrm b}) \cos \left(n\Delta\phi\right)
\end{eqnarray}
where the phase of $n^{\mathrm{th}}$ harmonic flow $\Psi_n$ (known as the event plane or EP) drops out in the convolution. 

The $v_n$ can be extracted directly by correlating single particles with the measured $n^{th}$ order EP (the event plane method) or via a Fourier transform of the two-particle azimuthal correlation (the two-particle correlation method). We present $v_n$ results from both methods based on $\sim8$~$\mu\mathrm{b}^{-1}$ Pb+Pb data from the 2011 LHC heavy ion run at $\sqrt{s_{\mathrm{NN}}}=2.76$~TeV~\cite{note}. A detailed comparison is made between the two methods, followed by an interpretation of the long-range structures in terms of geometric fluctuations and collective flow.
\section{Event Plane Method}
The event plane $\Psi_n$ is estimated using the transverse energy flow measured in towers from the Forward Calorimeter (FCal) within $3.3<|\eta|<4.8$. By default, each $v_n$ is measured by correlating tracks in the Inner Detector covering $|\eta|<2.5$ with the EP from the full FCal, followed by a resolution correction~\cite{Poskanzer:1998yz}. In a slightly modifed method (FCal$_{\mathrm{P(N)}}$), tracks are correlated with the EP calculated with FCal in the opposite hemisphere, i.e tracks with $\eta\ge0$ ($\eta<0$) are correlated with the EP obtained from FCal$_{\mathrm N}$ (FCal$_{\mathrm P}$).  The FCal$_{\mathrm{P(N)}}$ method greatly increases the pseudorapidity separation between the tracks and the EP from about 3 units to about 5 units, thus it is much less affected by the so-called ``non-flow'' correlations, which stem primarily from jet fragmentation and resonance decays. This is especially useful for measuring the pseudorapidity dependence of $v_n$. We present results from both methods, relying more on the full FCal method when better resolution is needed (e.g. for the higher-order harmonics).

\begin{figure}[b]
\centering
\includegraphics[width=0.4\columnwidth]{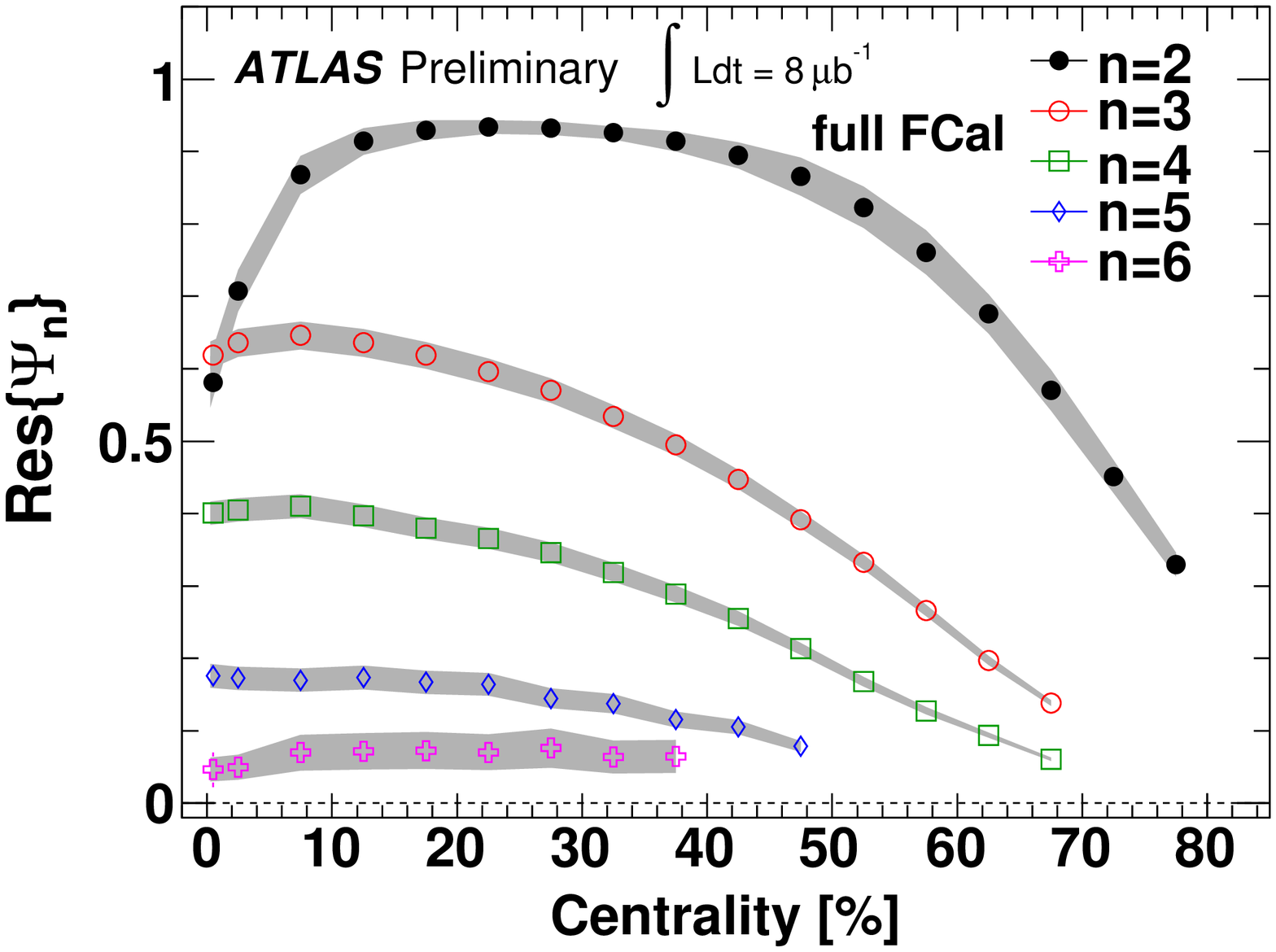}\includegraphics[width=0.4\columnwidth]{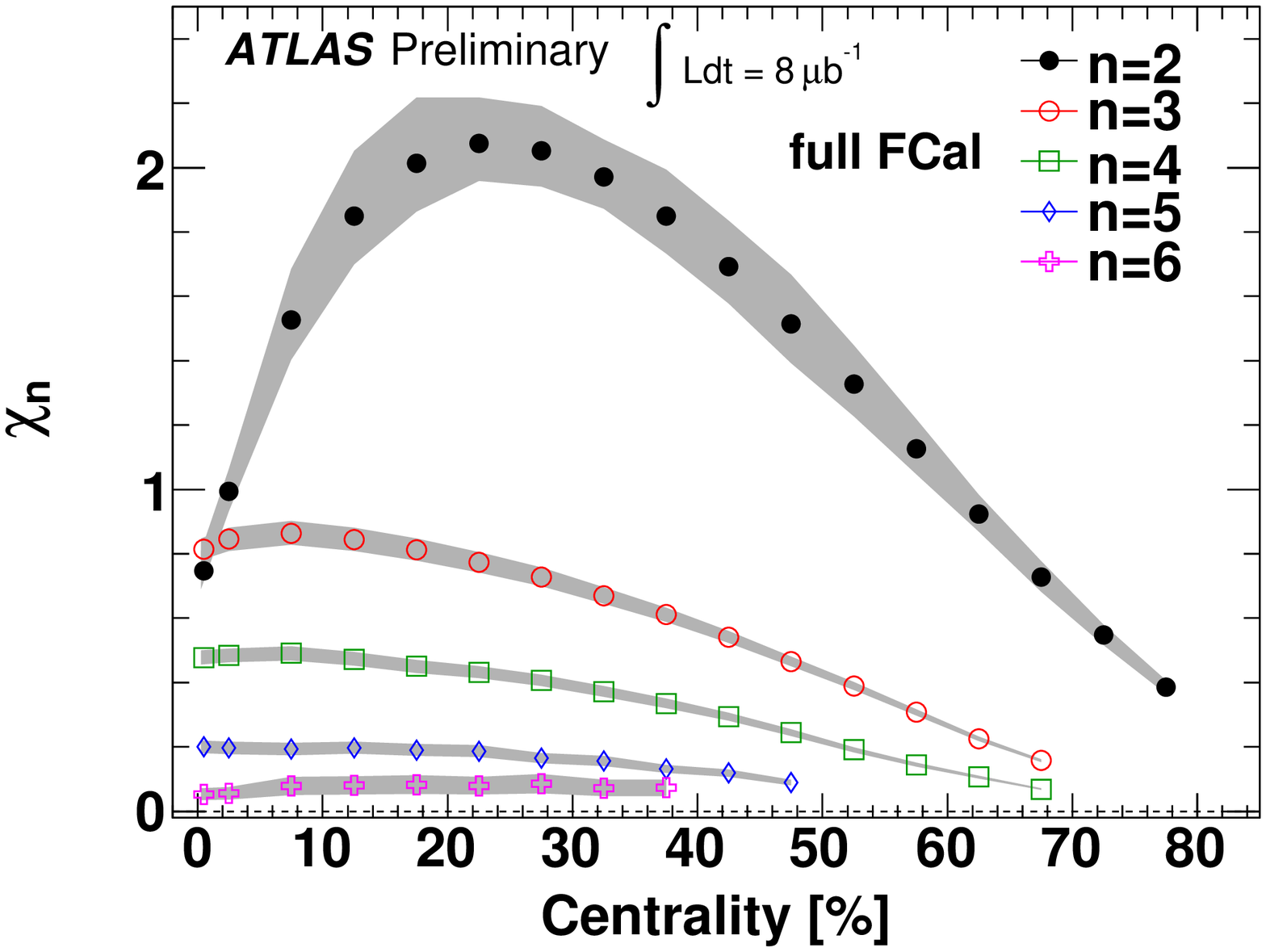}
\caption{The resolution (left) and resolution parameter $\chi_n$~\cite{Poskanzer:1998yz} (right) vs centrality (smaller number refer to more central collisions) for $n=2-6$ from the full FCal detector, together with their systematic uncertainty bands.}
\label{fig:a1}
\end{figure}

\begin{figure}[t]
\begin{center}
\includegraphics[width=0.3\linewidth]{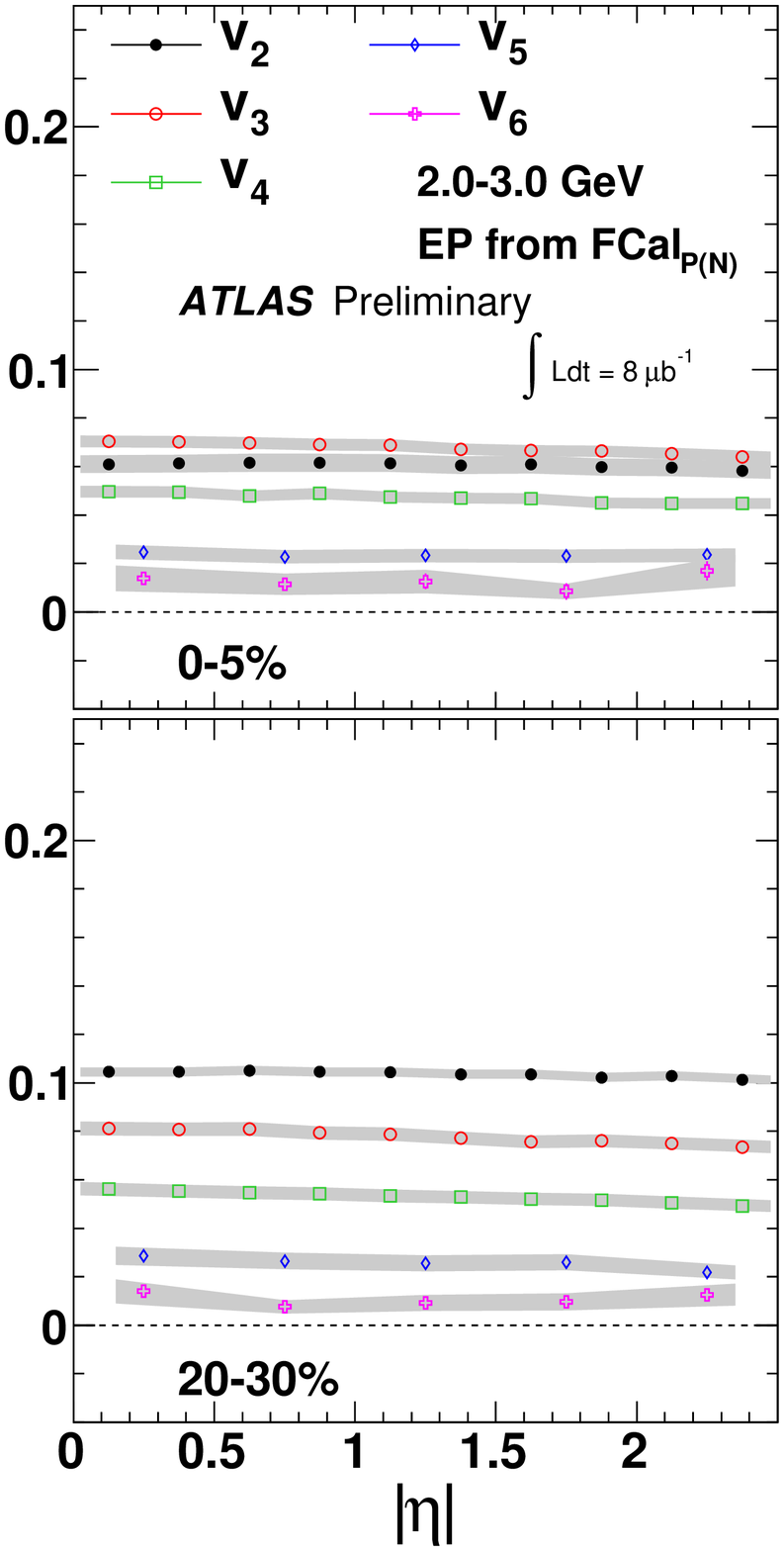}
\includegraphics[width=0.3\linewidth]{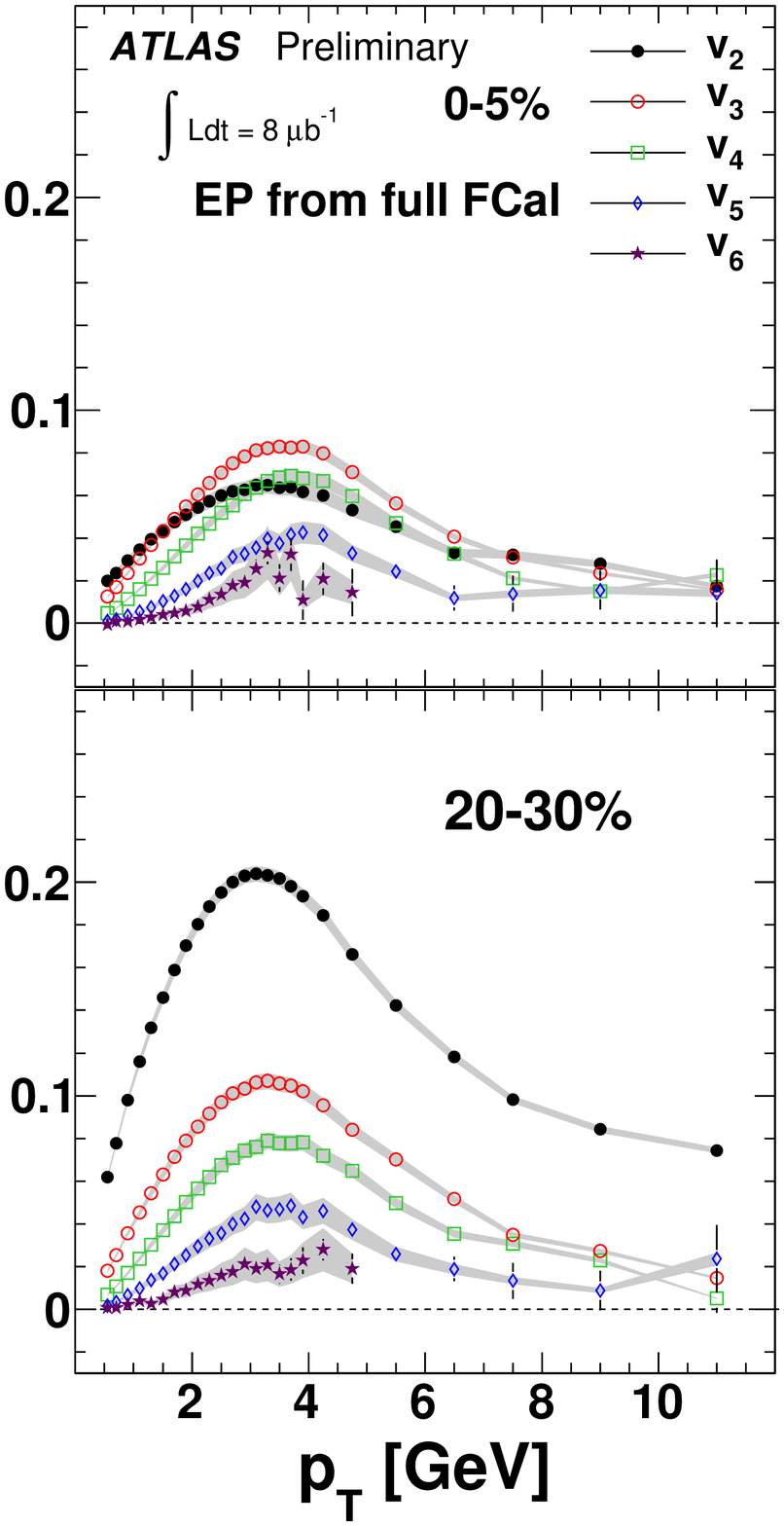}
\includegraphics[width=0.3\linewidth]{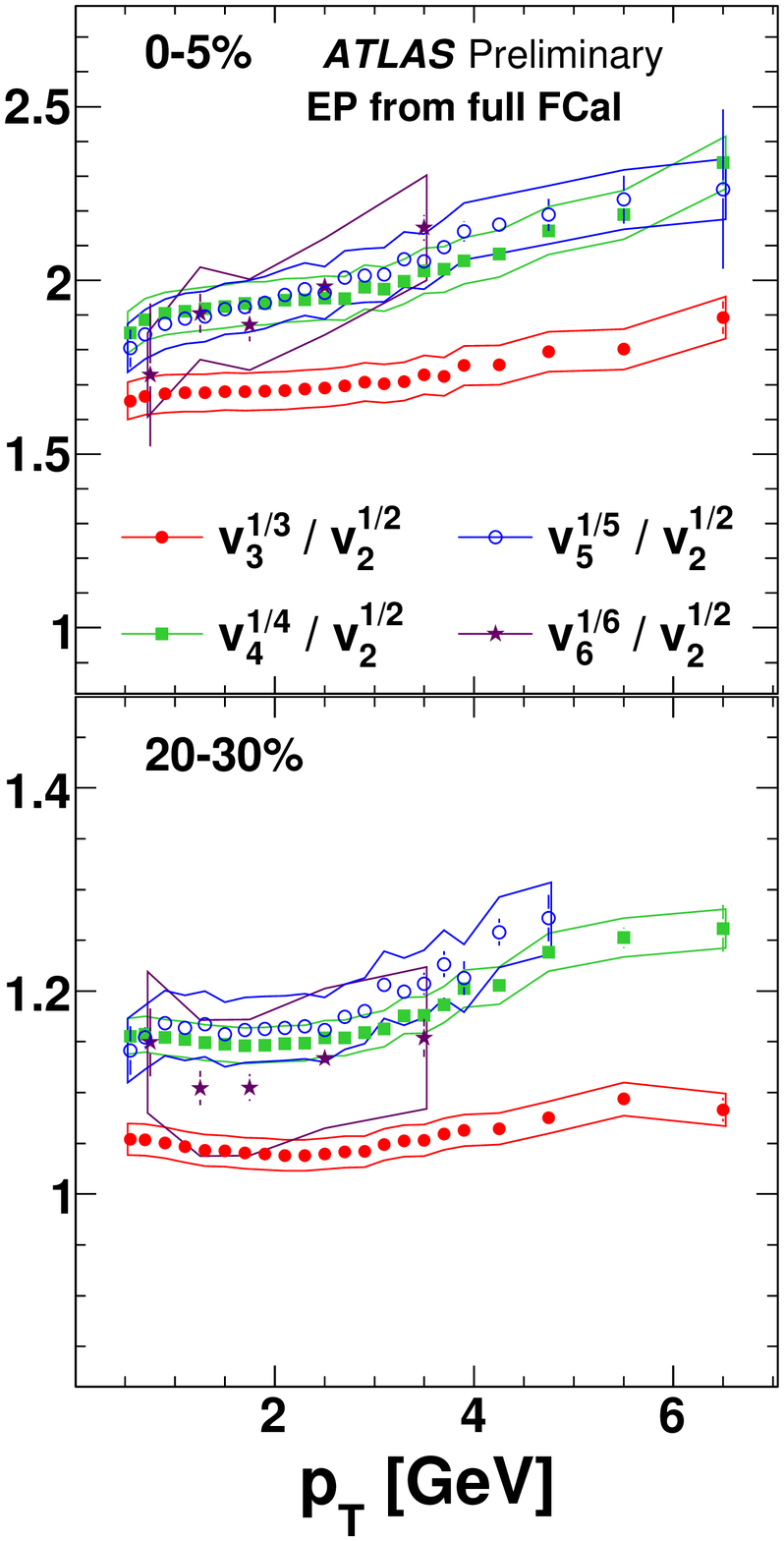}
\end{center}
\caption{(left) $\eta$ dependence of $v_n$ for $2<\pT<3$ GeV from FCal$_{\mathrm{P(N)}}$; (middle) $\pT$ dependence of $v_n$; (right) $\pT$ dependence of  $v_n^{1/n}/v_2^{1/2}$, for two centrality selections.}
\label{fig:a2} 
\end{figure}

Figure~\ref{fig:a1} shows the resolution factor in 5\% centrality intervals along with a 0-1\% most-central interval from the full FCal method. Significant values are observed for n=2-6. It is interesting to note that the resolution factor for $n=3$ exceeds that for $n=2$ in the 0-1\% most central collisions. 
Figure~\ref{fig:a2} summarizes the $\eta$ and $\pT$ dependence of $v_2-v_6$ for two centrality selections. A very weak $\eta$ dependence (less than 5\%) is observed for all $n$ within $|\eta|<2.5$.

All measured harmonics have a similar $\pT$ dependence: they first increase rapidly up to $\pT\sim3-4$ GeV and then fall. The similar $\pT$ dependence motivates us to find a simple scaling relation between different $v_n$: $v_n^{1/n} \propto v_2^{1/2}$. This is qualitatively understood in a "blast wave" scenario, where $v_n \propto \beta^n$, with $\beta$ being the radial flow velocity~\cite{Esumi:1997zz}. However, it should be pointed out that this scaling is only approximate. In particular, the ratios for $n=4-6$ are close to each other, while they are systematically higher than those for $n=3$.

\begin{figure}[h]
\begin{center}
\includegraphics[width=0.87\linewidth]{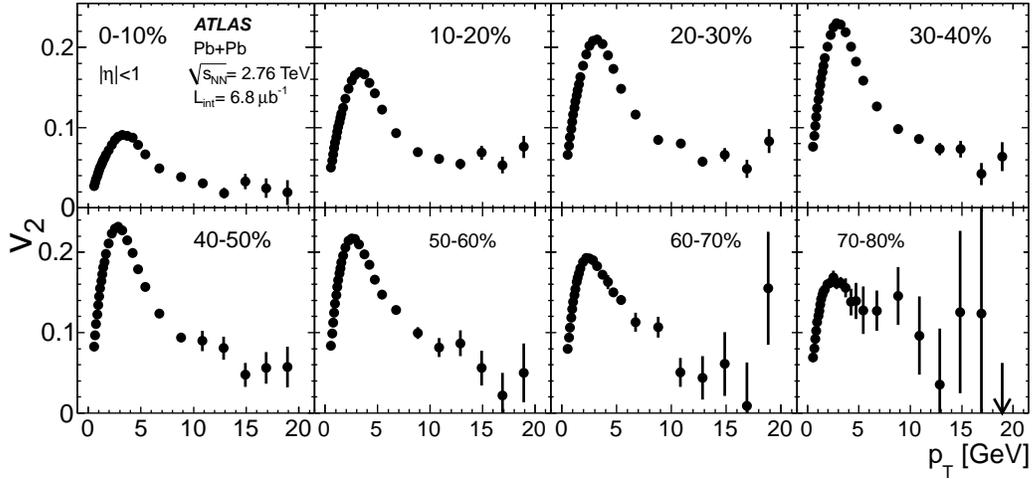}
\end{center}
\caption{$v_2$ vs $\pT$ at mid-rapidity for several centrality selections.}
\label{fig:a3}
\end{figure}

In Fig.~\ref{fig:a2}, $v_2-v_5$ are shown only up to 12 GeV, so the low $\pT$ region can be seen more clearly. However, ATLAS has measured $v_2$ out to much larger $\pT$, as shown in Figure~\ref{fig:a3}~\cite{adam}, with much smaller statistical errors at high $\pT$ than previous RHIC results~\cite{Adare:2010sp}. The $v_2$ clearly continue to drop out to 10-12 GeV, and only vary slowly beyond that, but at a level that is consistent with pQCD calculations~\cite{will}.

\section{Two-particle Correlation Method}
\begin{figure}[h]
\begin{center}
\includegraphics[width=0.33\linewidth]{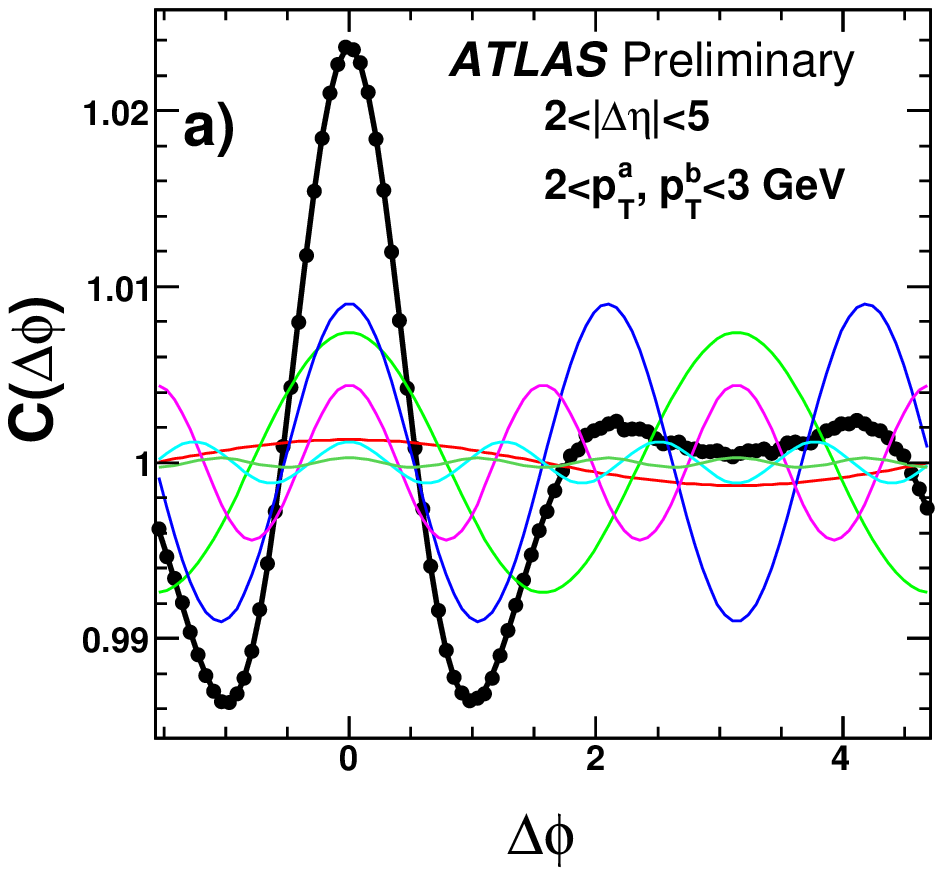}\includegraphics[width=0.32\linewidth]{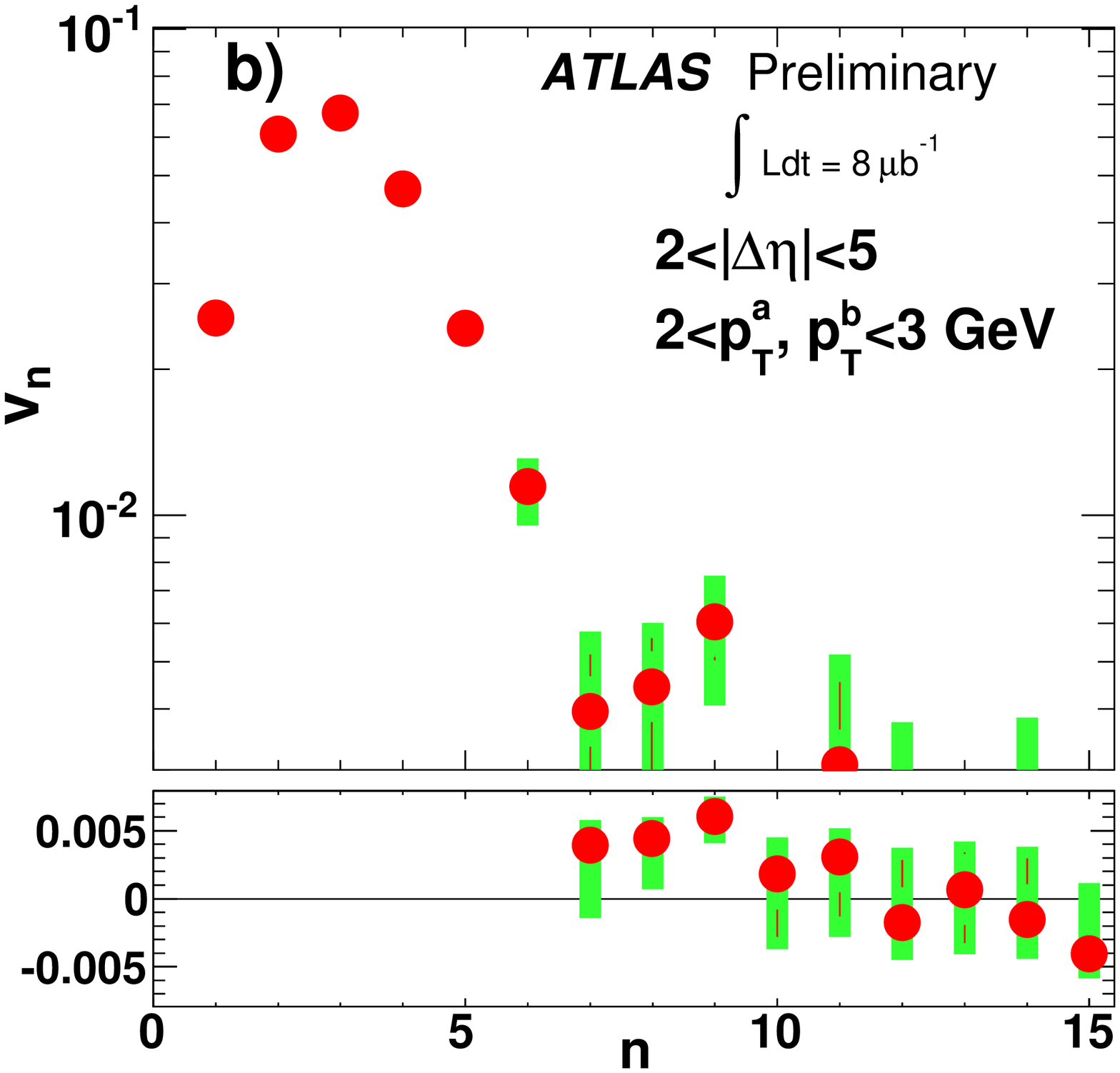}\includegraphics[width=0.32\linewidth]{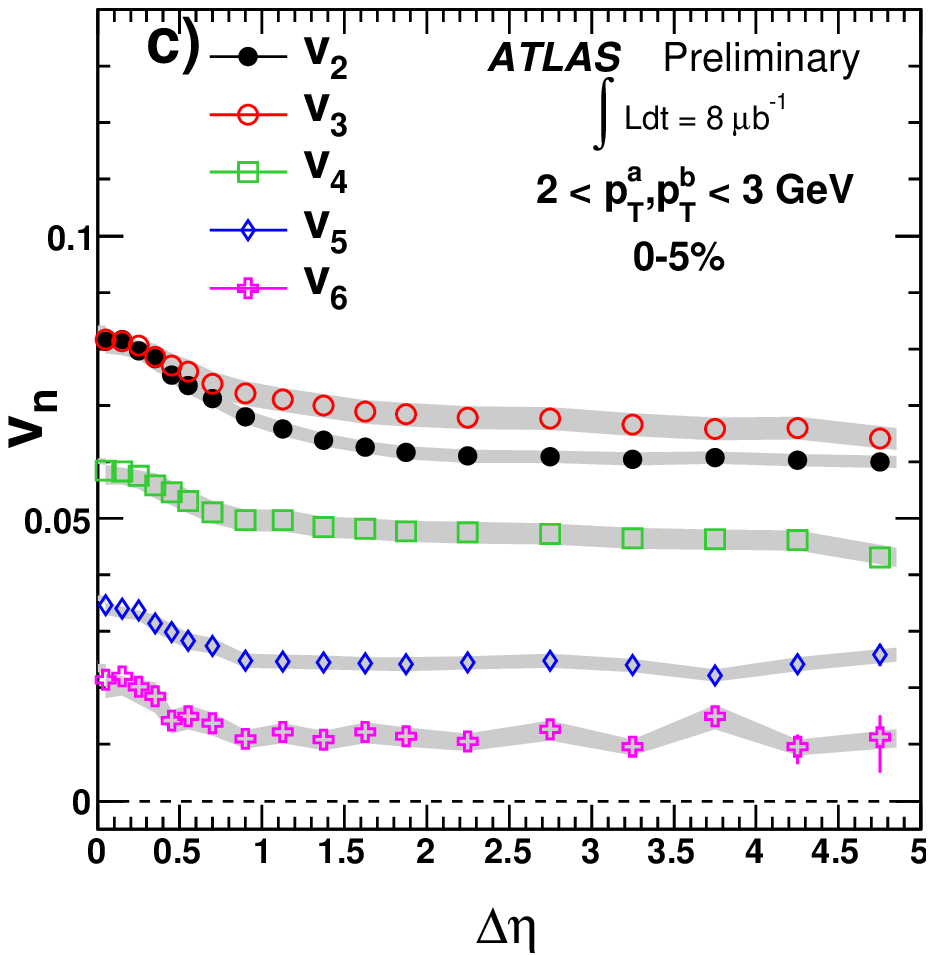}
\end{center}
\vspace*{-0.4cm}
\caption{\label{fig:b1} The steps involved in the extraction of the $v_n$ (2-3 GeV fixed-$\pT$ correlation in 0-5\% centrality): a) $\Delta\phi$ correlation function for $2<|\Delta\eta|<5$, overlaid with contributions from individual Fourier components and the sum, b) Fourier coefficient $v_n$ vs $n$, and c) $v_2-v_6$ vs $\Delta\eta$.}
\end{figure}

The correlation function is constructed by dividing the same-event pairs with mixed-event pair, with a pair acceptance extending to $|\Delta \eta| = 5$. The normalization is fixed by equating the counts of the same-event and mixed-event pairs in $2<|\Delta\eta|<5$, which is then applied for all $\Delta\eta$ slices. Each 1-D correlation function (obtained by integrating a selected $|\Delta\eta|$ range) is expanded into a Fourier series, with the coefficient $v_{n,n}$ calculated directly via a discrete Fourier transformation (DFT): $v_{n,n}= \langle\cos\left(n\Delta\phi\right)\rangle$. Figure~\ref{fig:b1}a shows one such projection for $2<|\Delta\eta|<5$ and the corresponding contributions from individual harmonic components.

If the observed modulations are due only to collective flow, then we expect $v_{n,n}$ to be factorizable into the product of two single-particle flow coefficients: 
\begin{eqnarray}
\label{eq:fac}
v_{n,n}(\pT^{\mathrm a},\pT^{\mathrm b}) = v_n(\pT^{\mathrm a})v_n(\pT^{\mathrm b})
\end{eqnarray}
Thus for correlations where the two particles are selected from the same $\pT$ range, we calculate the single-particle harmonic coefficient as $v_n=\sqrt{v_{n,n}}$. One such example is shown in Figure~\ref{fig:b1}b. We have repeated this procedure for each $\Delta\eta$ slice and the results are shown in Figure~\ref{fig:b1}c. The peak at small $\Delta\eta$, which comes from near-side jet fragmentation and resonance decay, is excluded by requiring $2<|\Delta\eta|<5$. 

\begin{figure}[t]
\includegraphics[width=1\linewidth]{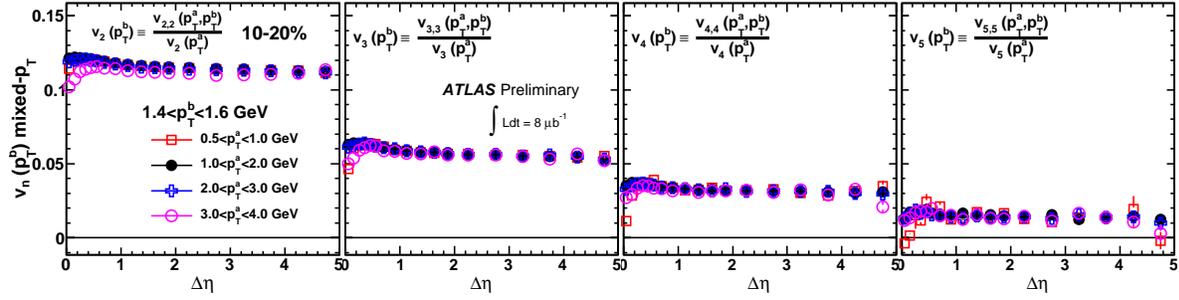}
\caption{$v_{n}(\pT^{\mathrm b})=\frac{v_{n,n}(\pT^{\mathrm a},\pT^{\mathrm b})}{v_n(\pT^{\mathrm a})}$ vs $\Delta\eta$ for target $\pT$ of $1.4<\pT^{\mathrm b}<1.6$ GeV, which are extracted from four reference $\pT^{\mathrm a}$ bins, supporting the factorization relation Eq.~\ref{eq:fac}.
}
\label{fig:b2}
\end{figure}

In order to confirm that the $v_n$ coefficients from the two-particle correlation method reflect collective flow, the factorization relation in Eq.~\ref{eq:fac} has been explicitly checked with correlations of pairs of tracks from different $\pT$ ranges. For correlations with $|\Delta\eta|>2$, this factorization indeed holds for $v_2-v_6$ at 5\%-10\% level for $\pT<3-4$ GeV in the $70\%$ most central events, where away-side jet contributions are small. One example of such explicit check is shown in Figure~\ref{fig:b2}. Conversely, the factorization is found to break down for $v_1$, which could be due to the fact that $v_1$ from collective flow changes sign going from negative $\eta$ to positive $\eta$~\cite{Abelev:2008jga}. However, similar effects can also be caused by momentum conservation effects, e.g. the recoil of the away-side jet in a di-jet system.

\begin{figure}[b]
\centering
\includegraphics[width=0.4\linewidth]{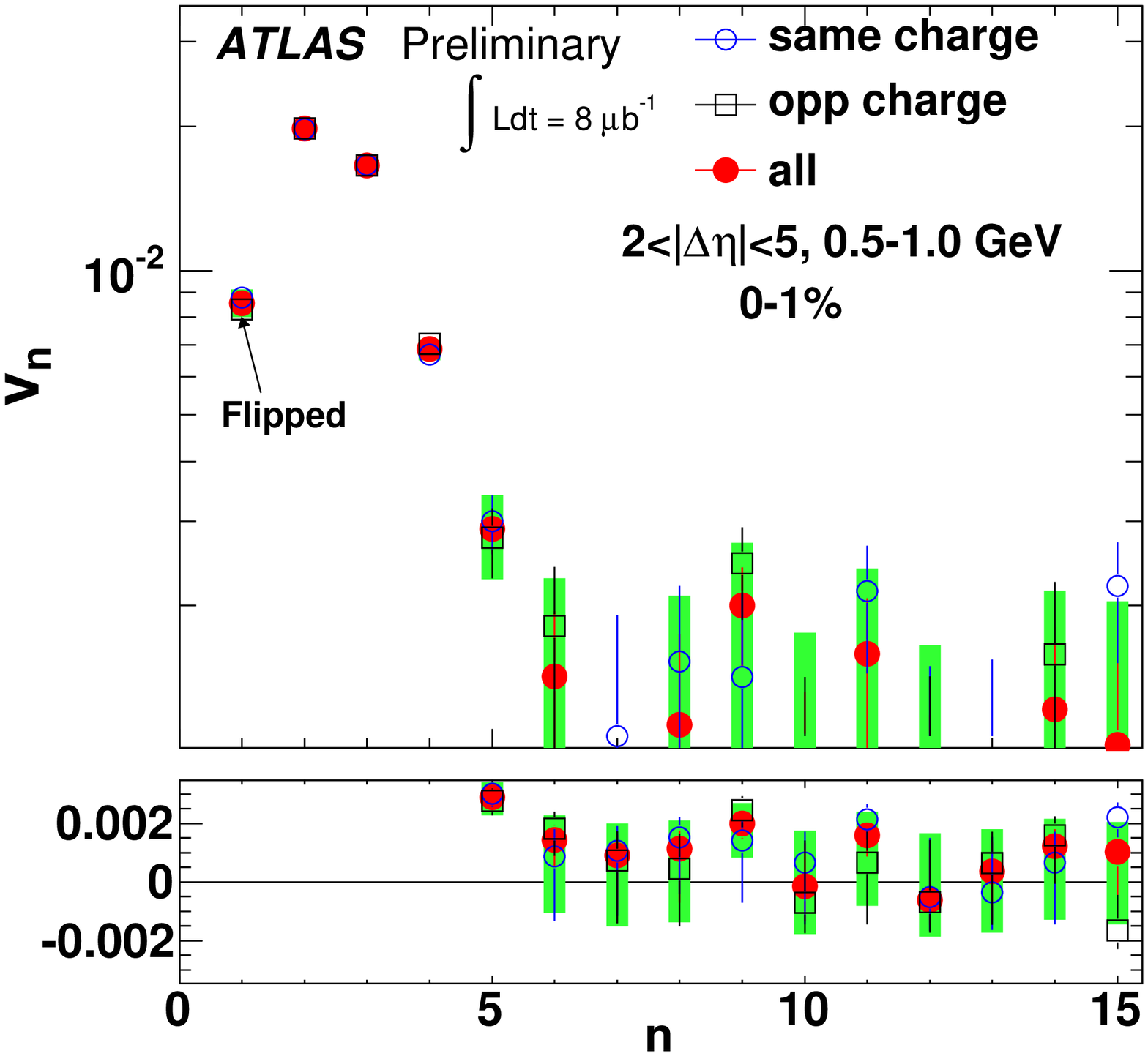}\includegraphics[width=0.4\linewidth]{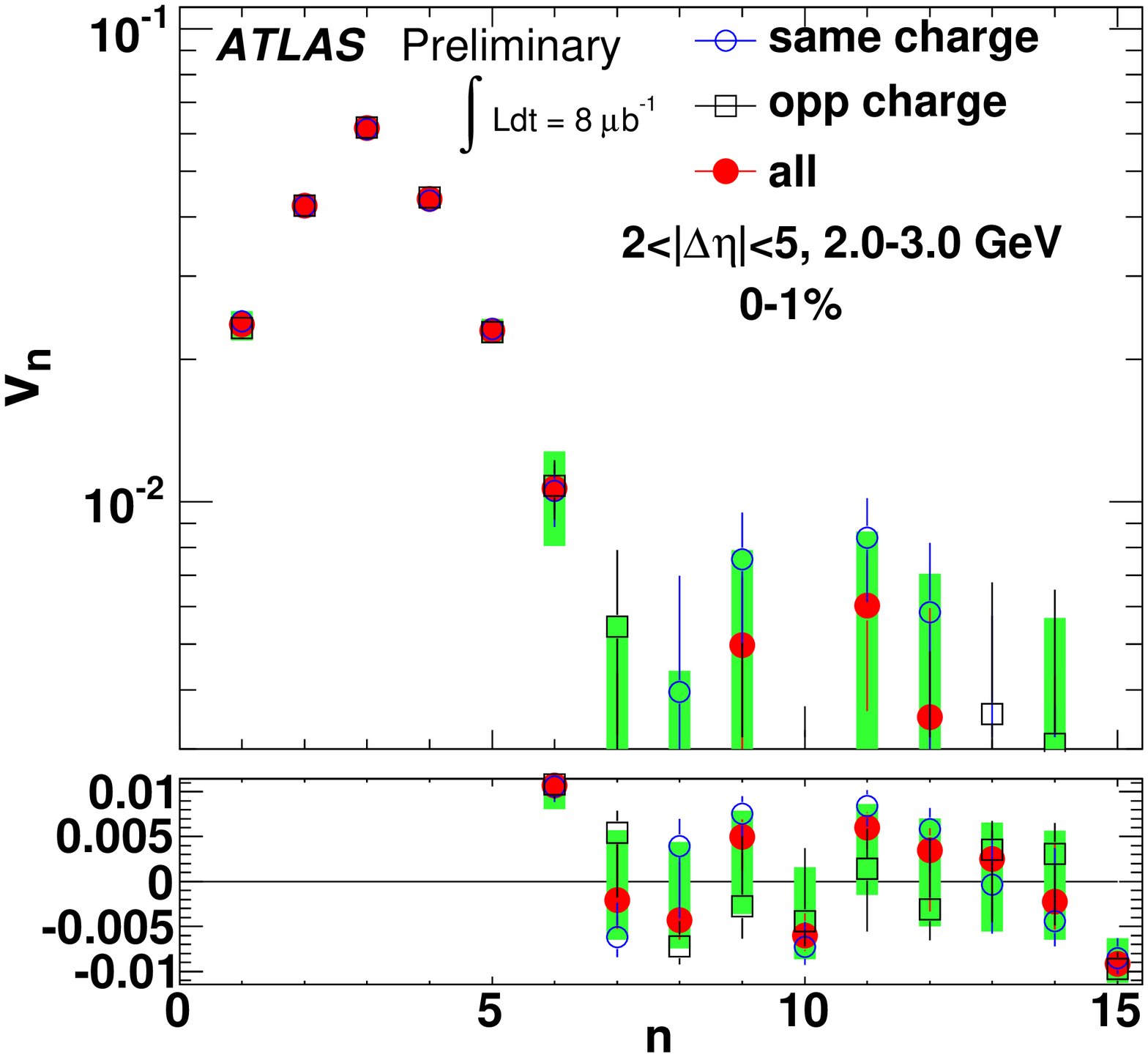}
\caption{$v_n$ vs $n$ from same-charge, opposite-charge and all pairs in 0-1\% centrality bin for two fixed-$\pT$ correlations: 0.5-1, 2-3 GeV.}
\label{fig:b3}
\end{figure}

Figures~\ref{fig:b3} shows the extracted Fourier spectra for two fixed-$\pT$ correlations with a large $\Delta\eta$ gap ($2<|\Delta\eta|<5$) in 0-1\%. $v_2$ is the largest at low $p_T$, but it becomes less than $v_3$ and $v_4$ for tracks in $2<\pT<3$ GeV. The rate with which the $v_n$ decrease with $n$ were shown to provide important insight on the acoustic horizon and viscous effects~\cite{Staig:2011as}.

\section{Comparison of the Methods and Discussion}
\begin{figure}[t]
\centering
\includegraphics[width=0.8\linewidth]{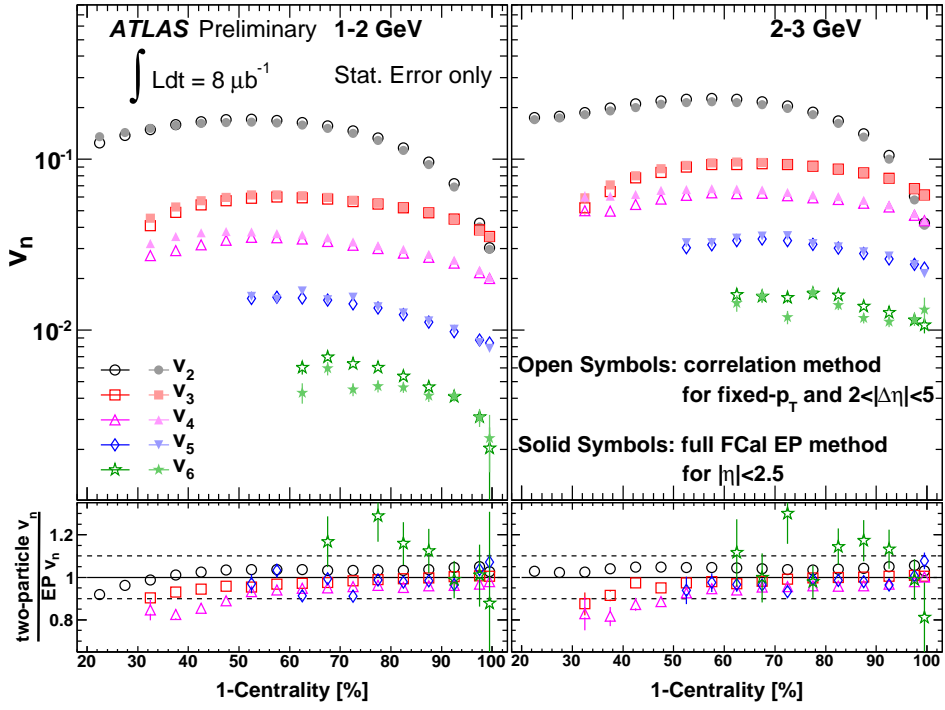}
\caption{Comparison of $v_n$ obtained from the two-particle fixed-$\pT$ correlation method (solid symbols) to the full FCal results (open symbols) for 1-2 GeV (left panels) and 2-3 GeV (right panels). The ratios between the two methods are shown in the bottom panels.}
\label{fig:c1}
\end{figure}

Figure~\ref{fig:c1} compares the centrality dependence of $v_n$ obtained from the two-particle correlation method and those from the EP method for two $\pT$ intervals. The agreements is within 5\% for $v_2-v_4$ over a broad centrality range, but worsen to about 10\% for $v_5$ and 15\% for $v_6$. However, they are within the quoted systematic errors shown in previous sections. 

\begin{figure}[h]
\centering
\includegraphics[width=0.9\linewidth]{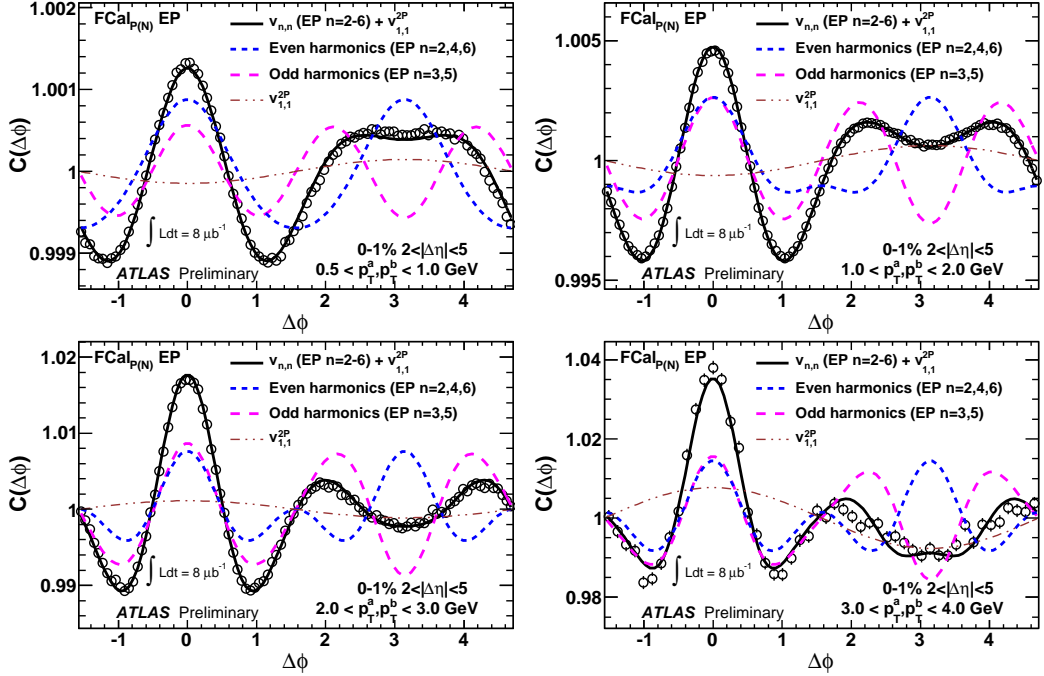}
\caption{Correlation function data compared with that reconstructed from $v_{1,1}$ from two-particle correlation and $v_{2}-v_6$ measured from EP method in 0-1\% centrality for different $\pT$ ranges.}
\label{fig:c2}
\end{figure}

The consistency between the two methods implies that the structure of the two-particle correlation at large $\Delta\eta$ can be largely accounted for by the collective flow. We check this explicitly by reconstructing the correlation function from $v_n$ measured by the EP method as:
\begin{eqnarray}
\label{eq:reco}
C(\Delta\phi) = b^{\mathrm{2p}}(1+2v_{1,1}^{\mathrm{2p}}\cos\Delta\phi+2\sum_{n=2}^{6}v_{n}^{\mathrm{EP,a}}v_{n}^{\mathrm{EP,b}}\cos n\Delta\phi)
\end{eqnarray}
where the $b^{\mathrm {2p}}$ and $v_{1,1}^{\mathrm {2p}}$ are the pedestal and first harmonic term from the two-particle correlation analysis, while the remaining terms are calculated from the $v_n$ measured using the EP method. Figure~\ref{fig:c2} shows excellent agreement between the measured and reconstructed correlation functions. This is more striking for the 2-3 and 3-4 GeV bin in the 0-1\% most central collisions, where even a more pronounced ``double hump'' structure can be well reproduced. The $v_{1,1}$ term not measured by the EP method, also plays a significant role, but it is not sufficient to explain the near and away-side structures obtained in the correlation function. Figure~\ref{fig:c2} also shows that the away-side ``double-hump'' is the result of the detailed interplay between the odd harmonics ( $v_3$ and $v_5$) and even harmonics ($v_2, v_4$, and $v_6$).

\begin{figure}[t]
\centering 
\includegraphics[width=0.9\linewidth]{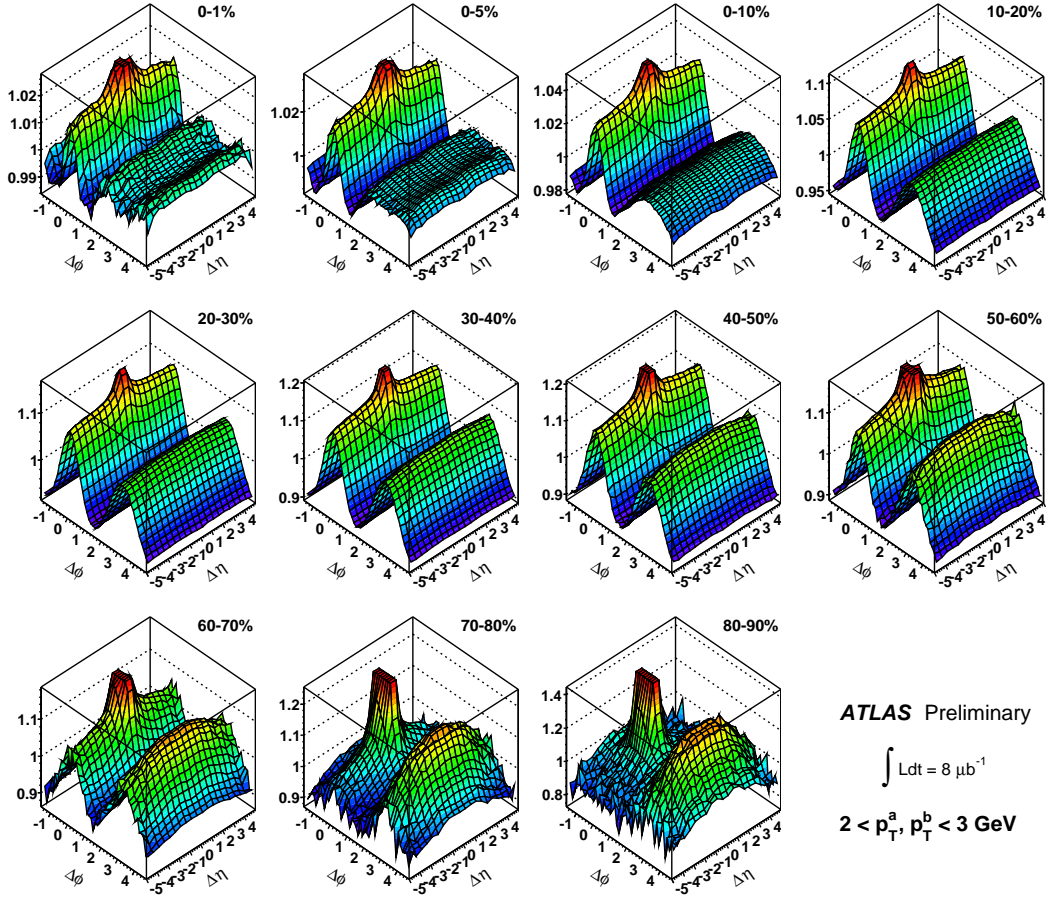}
\caption{2-D correlations for $2<\pT^{\mathrm a},\, \pT^{\mathrm b}<3$ GeV as a function of centrality. The near-side jet peak is truncated to better reveal long range structures.}
\label{fig:c3}
\end{figure}

\begin{figure}[t]
\centering
\includegraphics[width=1\linewidth]{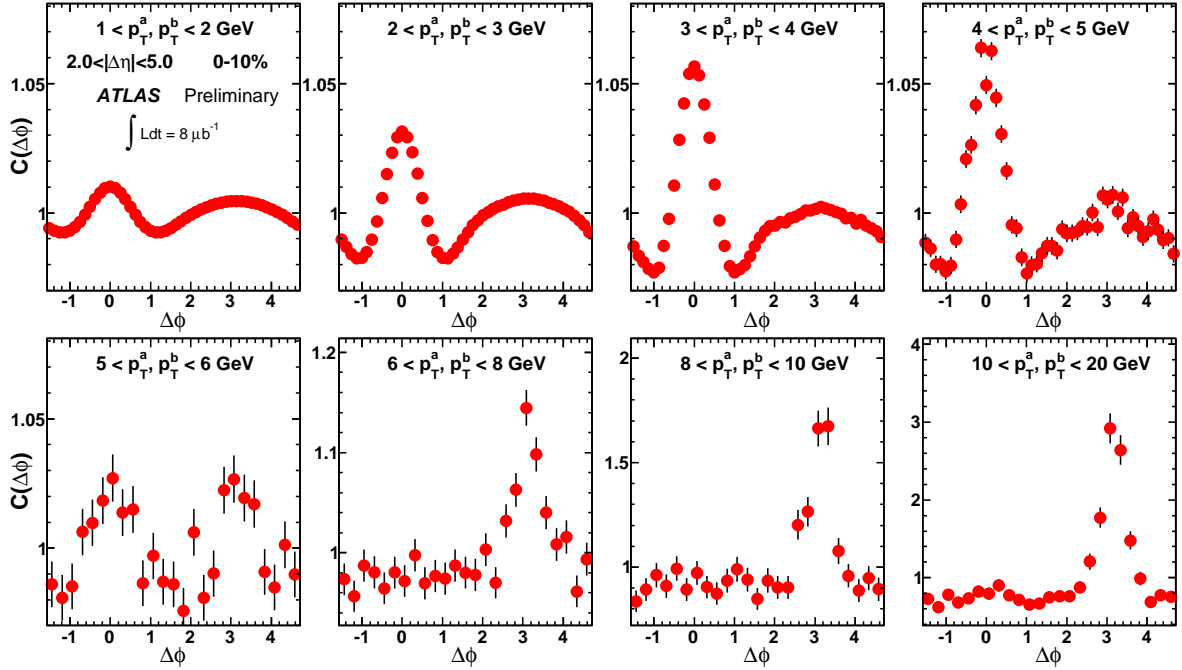}
\caption{$\pT$ evolution of two-particle $\Delta\phi$ fixed-$\pT$ correlations for 0-10\% centrality selection, with a large rapidity gap ($|\Delta\eta|>2$) to suppress the near-side jets and select only the long range components.}
\label{fig:c4}
\end{figure}

If the low $\pT$ correlation function with large $\Delta\eta$ gap is dominated by harmonic flow, an important question remains as to where in phase space the jet contributions can be observed. Figure~\ref{fig:c3} shows the centrality evolution of the 2-D correlation function for particles with $2<\pT<3$ GeV. While central events show structures that are long range in $\Delta\eta$ as well as a shorter-range jet correlation around $\Delta\eta=\Delta\phi=0$, moving towards more peripheral events show that these long-range structures eventually disappear, with clear jet-related peaks emerging on the away-side. Figure~\ref{fig:c4} shows the $\pT$ evolution of $\Delta\phi$ correlations in the 0-10\% most central collisions. A rapidity gap of $|\Delta\eta|>2$ is required such that the near-side peak reflects mainly the ``ridge'' contribution. Its magnitude first increases with $\pT$ to 4-5 GeV then decreases, reflecting the fact that all of the $v_n$ reach a maximum at 3-4 GeV as shown in Fig.~\ref{fig:a2}. A narrow away-side peak emerges in the 6-8 GeV bin, which quickly dominates the correlation structure at higher $\pT$. This away-side peak presumably comes from the fragmentation of the recoil jet. This figure qualitatively suggests that the transition in $\pT$ from being flow-dominated to jet-dominated, for correlations of tracks from the same $\pT$ range, happens somewhere around 6-8 GeV. Interestingly, this is the $\pT$ region where the single hadron suppression is the strongest~\cite{Aamodt:2010jd}.

\section{Conclusion}
In summary, the higher-order harmonic coefficients $v_2-v_6$ have been extracted both by correlating tracks with the event plane determined at forward rapidity and by using the two-particle correlation method with a large pseudorapidity gap ($|\Delta\eta|>2$). Significant $v_2-v_6$ are observed and they are consistent between the two methods for $\pT<4$ GeV. The $v_2-v_6$ are found to decrease only slightly with $|\eta|$ in $0<|\eta|<2.5$. All $v_n$ exhibit similar $\pT$ dependence, namely, they all increase with $\pT$ to around 3-4 GeV and then drop for higher $\pT$. However, the higher-order harmonics show a stronger $\pT$ variation, which is found to follow a simple scaling relation, $v_n^{1/n}\propto v_2^{1/2}$, varying only weakly with $\pT$. We find the main structures of the two-particle correlation at $|\Delta\eta|>2$ can be explained by $v_{2}-v_{6}$ which largely reflect collective flow, and a $v_{1}$ term which accounts for momentum conservation effects (possibly including a global directed flow). We conclude that the low $\pT$ correlation functions do not allow significant contributions from a short $\Delta\eta$ range medium response to jets. Fluctuations in the initial state geometry, along with a non-zero viscosity of the medium, are potentially responsible for the detailed behavior of the harmonic coefficients. A detailed comparison with viscous hydrodynamic calculations~\cite{Qin:2010pf} can help elucidate the nature of these fluctuations and better constrain the transport properties of the hot, dense medium.


\section*{References}
\begin{thebibliography}{10}
\bibitem{Adare:2008cqb}
  B.~I.~Abelev {\it et al.}
  Phys.\ Rev.\  C {\bf80}, 064912 (2009);
  A.~Adare {\it et al.},
  Phys.\ Rev.\  C {\bf 78}, 014901 (2008);
\bibitem{CasalderreySolana:2004qm}
  J.~Casalderrey-Solana, E.~V.~Shuryak, D.~Teaney,
  J.\ Phys.\ Conf.\ Ser.\  {\bf 27}, 22-31 (2005).
\bibitem{Alver:2010gr}
  B.~Alver and G.~Roland,
  Phys.\ Rev.\  C {\bf 81}, 054905 (2010)
  [Erratum-ibid.\  C {\bf 82}, 039903 (2010)].
\bibitem{Poskanzer:1998yz}
  A.~M.~Poskanzer and S.~A.~Voloshin,
  Phys.\ Rev.\  C {\bf 58}, 1671 (1998).
\bibitem{note}
ATLAS Collaboration, Measurement of elliptic flow and higher-order flow coefficients with the ATLAS detector in $\sqrt{s_{\rm NN}}$=2.76 TeV Pb+Pb collisions, ATLAS-CONF-2011-074, May 2011. \url{http://cdsweb.cern.ch/record/1352458}.
\bibitem{Esumi:1997zz}
  S.~Esumi, S.~Chapman, H.~van Hecke and N.~Xu,
  Phys.\ Rev.\  C {\bf 55}, R2163 (1997).
\bibitem{adam}
A.~Trzupek, these proceedings.
\bibitem{Adare:2010sp}
  A.~Adare {\it et al.},
  Phys.\ Rev.\ Lett.\  {\bf 105}, 142301 (2010).
\bibitem{will}
W.~A.~Horowitz, these proceedings.
\bibitem{Abelev:2008jga}
B.~I.~Abelev {\it et al.}  Phys.\ Rev.\ Lett.\  {\bf 101}, 252301 (2008).
\bibitem{Staig:2011as}
  P.~Staig and E.~Shuryak,
    [arXiv:1106.3243 [hep-ph]].
\bibitem{Aamodt:2010jd}
K.~Aamodt {\it et al.}
  Phys.\ Lett.\  B {\bf 696}, 30 (2011).

\bibitem{Qin:2010pf}
  G.~Y.~Qin, H.~Petersen, S.~A.~Bass and B.~Muller,
  Phys.\ Rev.\  C {\bf 82}, 064903 (2010);  B.~Schenke, S.~Jeon, C.~Gale,  [arXiv:1102.0575 [hep-ph]];
  Z.~Qiu and U.~W.~Heinz,  arXiv:1104.0650 [nucl-th].
\end{thebibliography}
\end{document}